\documentclass{elsart}
\usepackage{amssymb}
\usepackage{graphicx}

\begin{document}

\begin{frontmatter}

\title{The {\em Invar} tensor package: \\ Differential invariants of Riemann}
\author[a]{J.M. Mart\'{\i}n-Garc\'{\i}a}%
\author[a]{, D. Yllanes}%
\address[a]{Instituto de Estructura de la Materia, CSIC, \\
C/ Serrano 123, Madrid 28006, Spain}
\author[b,c]{R. Portugal}%

\address[b]{Department of Applied Mathematics, University of Waterloo, \\
Waterloo, Ontario, Canada}

\address[c]{Laborat\'orio Nacional de Computa\c{c}\~ao Cient\'{\i}fica
(LNCC),\\ Av. Get\'ulio Vargas 333, Petr\'opolis, RJ, CEP 25651-075,
Brazil}

\begin{abstract}
The long standing problem of the relations among the scalar
invariants of the Riemann tensor is computationally solved for all
$6\cdot 10^{23}$ objects with up to 12 derivatives of the metric.
This covers cases ranging from products of up to 6 undifferentiated
Riemann tensors to cases with up to 10 covariant derivatives of a
single Riemann. We extend our computer algebra system
{\em Invar} to produce within seconds a canonical form for
any of those objects in terms of a basis. The process is as follows:
(1) an invariant is converted in real time into a canonical form with
respect to the permutation symmetries of the Riemann tensor;
(2) {\em Invar} reads a database of more than $6\cdot 10^5$ relations
and applies those coming from the cyclic symmetry of the Riemann tensor;
(3) then applies the relations coming from the Bianchi identity,
(4) the relations coming from commutations of covariant derivatives,
(5) the dimensionally-dependent identities for dimension 4, and finally
(6) simplifies invariants that can be expressed as product of dual
invariants. {\em Invar} runs on top of the tensor computer algebra systems
{\em xTensor} (for {\em Mathematica}) and {\em Canon} (for {\em
Maple}).

\end{abstract}


\end{frontmatter}

\newpage
{\footnotesize
\section*{Program summary}
\textit{Title of program:} Invar Tensor Package v. 2.0
\\
\textit{Catalogue identifier:}
\\
\textit{Program obtainable from:} (submitted to Computer Physics Communications) \\
\mbox{}\qquad \texttt{http://www.lncc.br/$\sim$portugal/Invar.html} (Maple version) and \\
\mbox{}\qquad \texttt{http://metric.iem.csic.es/Martin-Garcia/xAct/Invar/} (Mathematica)
\\
\textit{Reference in CPC to previous version:} Computer Physics
Communications 177 (2007) 640--648
\\
\textit{Catalogue identifier of previous version:} ADSP
\\
\textit{Does the new version supersede the original program?:} Yes.
The previous version (1.0) only handled algebraic invariants.
The current version (2.0) has been extended to cover differential
invariants as well.
\\
\textit{Computers:} Any computer running Mathematica versions 5.0
to 6.0 or Maple versions 9 to 11
\\
\textit{Operating systems under which the new version has been
tested:} Linux, Unix, Windows XP, MacOS
\\
\textit{Programming language:} Mathematica and Maple
\\
\textit{Memory required to execute with typical data:} 100 Mb
\\
\textit{No. of bits in a word:} 64 or 32
\\
\textit{No. of processors used:} 1
\\
\textit{No. of bytes in distributed program, including test data,
etc.:} Code $<$ 1Mb; database: 40 Mb; 13 expanded cases order 12 at
commutation step: 250 Mb.
\\
\textit{Distribution format:} Unencoded compressed tar file
\\
\textit{Nature of physical problem:} Manipulation and
simplification of scalar polynomial expressions formed from the Riemann
tensor and its covariant derivatives.
\\
\textit{Method of solution:} Algorithms of computational group
theory to simplify expressions with tensors that obey permutation
symmetries. Tables of syzygies of the scalar invariants of the
Riemann tensor.
\\
\textit{Restrictions on the complexity of the problem:} The
present version only handles scalars, but not expressions with
free indices.
\\
\textit{Typical running time:} One second to fully reduce
any monomial of the Riemann tensor up to degree 7 or order 10
in terms of independent invariants.
}

\section{Introduction}
Extracting information from the Riemann curvature tensor of a metric
field on a manifold is not an easy task due to its complicated
algebraic structure. The problem can be handled when restricting to
the algebraic invariants of the Riemann tensor, for which the
problem of giving all of them in terms of a basis has been recently
solved, after decades of continued work \cite{Invar, Sneddon,
LimCarminati}. However, this is frequently not enough and the
differential invariants of Riemann are then required. For example,
Bonnor \cite{Bonnor} has shown that the acceleration in the `photon
rocket' metric does not affect the algebraic Riemann invariants, but
changes the differential invariants. In this way it is possible to
introduce singularities in the curvature which are unnoticed by the
algebraic curvature invariants \cite{MusgraveLake}. Differential
invariants of Riemann are also required to compute the different
loop-orders of renormalization of the Einstein-Hilbert Lagrangian
\cite{GoroffSagnotti}, or in dealing with its generalizations, like
the Lagrangian $L(R, \nabla^2 R, \ldots \nabla^{2n} R)$ proposed in
\cite{Wands}, or the general diffeomorphism invariant Lagrangian
$L(g_{ab}, R_{bcde}, \nabla_{a_1}R_{bcde}, \ldots,
\nabla_{(a_1}\cdots\nabla_{a_m)}R_{bcde}, \ldots)$ analyzed in
\cite{IyerWald}.

In spite of the large amount of effort dedicated to the algebraic
invariants, very little has been said about how to manipulate large
families of differential invariants, except for the important work
by Fulling {\em et al.} \cite{Fulling}, because the problem is much more
complicated. Here we shall apply and generalize the techniques we proposed
in \cite{Invar} to cover this case, extending our tensor computer algebra
system {\em Invar} to handle both the algebraic and the differential
invariants of the Riemann tensor.

\section{The problem}
As in \cite{Invar} we shall work on a manifold of dimension $d$
with a metric field $g_{ab}$
and its associated structures: the (torsionless) Levi-Civita
connection $\nabla_a$, its Riemann tensor $R_{abcd}$ and (when
restricting to $d=4$) the totally antisymmetric tensor
$\epsilon_{abcd}$, such that $\nabla_e g_{ab}=0$ and
$\nabla_e \epsilon_{abcd} = 0$.

Our main objective is constructing a basis of independent {\em invariants}
(that is, scalars formed from contraction of several Riemann tensors and
their $\nabla$-derivatives) and {\em dual invariants} (those also having an
$\epsilon_{abcd}$ tensor), along with all polynomial relations giving
any other {\em dependent} invariant in terms of those in the basis.
The restricted algebraic case without derivatives of Riemann has been
essentially solved, both from the computational point of view \cite{Invar}
and, after a long series of contributions, from the theoretical point
of view \cite{Sneddon,LimCarminati}. The general case is much more
complicated and remains, however, nearly unexplored. The only article
known to us in this direction is the work by Fulling {\em et al.}
\cite{Fulling} in which, using Young tableaux techniques, they were able
to compute numbers of independent {\em nondual} invariants in the basis for
a variety of cases, in different dimensions. They also gave the members
of the basis for the simpler of those cases, but no expansions of the
dependent invariants were provided, rendering the result unpractical.
This article fills in that important gap by recomputing the basis
of invariants (hence confirming for the first time the results in
\cite{Fulling}) together with a database of expressions of any other
independent invariant in terms of the basis.

\section{Notations}

Following Fulling {\em et al.\/} \cite{Fulling} we separate the set of
all monomial invariants of {\em degree} $n$ (the number of Riemann
tensors) in subsets ${\cal R}_{\{\lambda_1,\ldots,\lambda_n\}}$, where
$\lambda_i$ is the differentiation order of the $i$-th Riemann tensor,
assuming the tensors have been sorted such that
$\lambda_i \le \lambda_{i+1}$. An $n$-tuple
$\{\lambda_1,\ldots,\lambda_n\}$ will be referred to as a {\em case},
with the corresponding invariants having $N=4n+\sum_{i=1}^n \lambda_i$
indices and order $\Lambda = 2n+\sum_{i=1}^n \lambda_i$, the number of
derivatives of the metric, not to be confused with the total number of
derivatives of the Riemann tensors, which is $\sum_{i=1}^n \lambda_i$.
Both $N$ and $\Lambda$ are always even numbers because we only consider
scalar expressions, with all indices paired among them.
For example an invariant of the case $\{0,1,3\}$, hence with
$N=16$ indices and order $\Lambda=10$, is
\begin{equation} \label{invariant}
R_{abcd} \; \nabla_e R^{ecfg} \; \nabla^a\nabla_f\nabla_h R^{bdh}{}_g .
\end{equation}
Sets of dual invariants and dual cases will be denoted with an asterisk,
as in ${\cal R}^*_{\{\lambda_1,\ldots,\lambda_n\}}$ or
$\{\lambda_1,\ldots,\lambda_n\}^*$. The algebraic
cases considered in \cite{Invar} are denoted here as
$\{0,\stackrel{n}{\ldots},0\}$ for degree $n$.

We shall see that commutation of derivatives produces relations among
invariants of different cases, because it converts second derivatives
into additional Riemann tensors. This operation changes both the degree
$n$ and the differentiation orders $\lambda_i$ of the Riemann tensors,
but not the total order $\Lambda$ of metric derivatives.
This allows a simple classification of the relations, which
are all homogeneous in $\Lambda$. Following again ref. \cite{Fulling}
we give results up to $\Lambda=12$, which consistently fills the gap
among the algebraic cases of degrees 1 to 7 (that is $\Lambda=2$ to
$\Lambda=14$) in \cite{Invar}. Concerning duals, we give
results up to $\Lambda=8$, also consistent with the algebraic degrees
1 to 5 of \cite{Invar}. See tables \ref{table12345}, \ref{table67},
\ref{dualtable}
for the actual list of cases considered in this investigation, already
sorted by their corresponding values of $\Lambda$: there are 48 nondual
cases and 15 dual cases to treat. In the following, when talking about
total numbers of invariants and relations, we always include the 7
nondual and 5 dual algebraic cases already considered in \cite{Invar}.

\section{Algorithms}

Relations among Riemann invariants are a consequence of the
symmetries obeyed by the Riemann tensor and its $\nabla$-derivatives.
There are six different symmetries we can exploit and so we proceed in
six respective consecutive {\em steps}, each producing new relations
and therefore decreasing the number of independent invariants, as
shown in tables \ref{table12345}, \ref{table67} and \ref{dualtable}
(note that steps 3 and 4 are new with respect to \cite{Invar}):
\begin{enumerate}
\item Permutation symmetries. The Riemann tensor obeys the following
symmetries under permutations of indices:
\begin{equation}
R_{bacd} = - R_{abcd}, \qquad
R_{cdab} = R_{abcd} .
\end{equation}
As explained in \cite{Invar} this type of tensor symmetry, and the
induced symmetries in tensor products, can be efficiently handled
using fast algorithms for manipulation of permutation groups
\cite{Renato}. In our system, each monomial is converted in real
time to its canonical form. For example, the canonical form of
invariant (\ref{invariant}) is
\begin{equation}\label{canonical}
- R^{abcd} \, R_a{}^{efg}{}_{;e} \, R_{bcf}{}^h{}_{;hgd}
\equiv - I_{\{0,1,3\},2595} .
\end{equation}
This process is very fast: see Figure \ref{timings7Riemann} for
an histogram of timings of a case with 28 indices. In this way,
we enormously reduce the number of invariants we need to control.
There are $16!\sim 2\cdot 10^{13}$ invariants for case \{0,1,3\}.
As shown in Table \ref{table12345}, there are 3237 different canonical
forms (column `Canon'), and if we remove products of invariants of
lower degree, this number is further reduced to 3099 (column
`Invars'), which are indexed from $I_{\{0,1,3\},1}$ to
$I_{\{0,1,3\},3099}$.
The whole column `Invars' has been constructed by canonicalization
of a supercomplete set of more than 20 million permutations in 15 hours,
giving 640\,119 different nondual canonical forms and 7698 canonical duals.
Those canonical forms were then sorted within each case according
to the following priorities: i) invariants with more (differentiated)
Ricci scalars are sorted first; ii) invariants with more Ricci tensors
are sorted first; iii) more Laplacians first; iv) ordering based on
indices. The indexing of algebraic invariants in the first version
of {\em Invar} has been preserved for backwards compatibility.

We do not consider at this step symmetries coming from permutation
of covariant derivatives acting on scalar expressions. They will
be treated in step 4.

\begin{figure}[t!]
\begin{center}
\includegraphics[width=11cm]{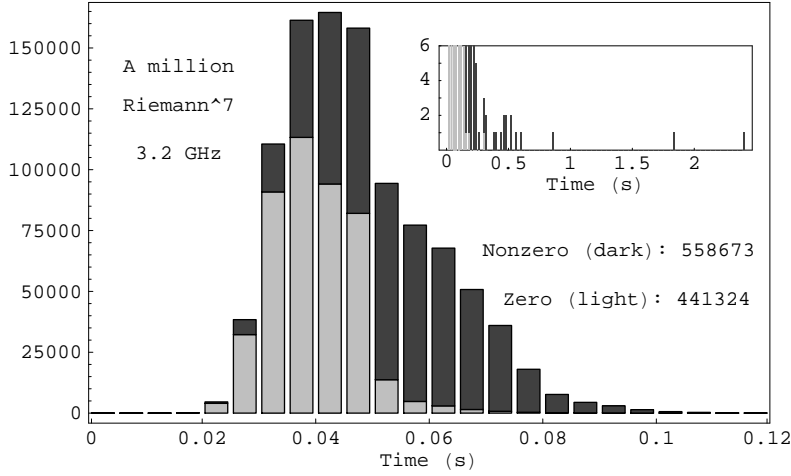}
\end{center}
\caption{\label{timings7Riemann}
Histogram of timings of canonicalization with {\em xTensor}
\cite{xTensor}
of a million algebraic invariants of degree 7. The average is below
0.05s and only two cases (with exceptionally high symmetry groups
involved) take more than a second.}
\end{figure}

\item Cyclic symmetry.
The Riemann tensor obeys the multiterm symmetry $R_{a[bcd]} = 0$. For
each canonical invariant after step 1 we generate several cyclic
relations by replacing $R_{abcd}$ by its 3-index antisymmetrized part
in all possible inequivalent ways. For example, from the canonical
form (\ref{canonical}) we get three cyclic relations (already
canonicalized):
\begin{eqnarray*}
R^{abcd}\, R_a{}^{efg}{}_{;e}\,
(2\, R_{bcf}{}^h{}_{;hgd} + R_{cdf}{}^h{}_{;hgb} )
&=& 0 , \\
R^{abcd}\, R_a{}^{efg}{}_{;e}\,
(R_{bcf}{}^h{}_{;hgd} - R_{bfc}{}^h{}_{;hgd} + R_b{}^h{}_{cf;hgd})
&=& 0 , \\
R^{abcd}\, (
  R_a{}^{efg}{}_{;e}\, R_{bcf}{}^h{}_{;hgd}
+ R_a{}^{efg}{}_{;f}\, R_{bcg}{}^h{}_{;hed}
- R_a{}^{efg}{}_{;f}\, R_{bce}{}^h{}_{;hgd} )
&=& 0 .
\end{eqnarray*}
\newpage
In terms of indexed invariants those are respectively
\begin{eqnarray*}
2 I_{\{0,1,3\},2595} + I_{\{0,1,3\},2610} &=& 0 , \\
I_{\{0,1,3\},2595} - I_{\{0,1,3\},2601} + I_{\{0,1,3\},2757} &=& 0 , \\
I_{\{0,1,3\},2595} - I_{\{0,1,3\},2634} + I_{\{0,1,3\},2640} &=& 0 ,
\end{eqnarray*}
allowing to express $I_{\{0,1,3\},2610}$, $I_{\{0,1,3\},2757}$ and
$I_{\{0,1,3\},2640}$ in terms of lower-index invariants.

\item Bianchi identity.
The Riemann tensor obeys the multiterm symmetry $R_{ab[cd;e]}=0$.
(Note that this, together with the previous permutation and cyclic
symmetries, make the tensor $R_{abcd;e}$ vanish under
antisymmetrization of any set of 3, 4 or 5 indices.) Again, from
every one of the independent invariants after step 1, two equations
are constructed by replacing $\nabla_a R_{bcde}$ by $\nabla_{[a}
R_{bc]de}$ and $\nabla_{[a} R_{de]bc}$. The results are then
simplified using the information of steps 1 and 2. For example from
the canonical invariant (\ref{canonical}) we get a zero equation and
\begin{displaymath}
R^{abcd}(
  R_a{}^{efg}{}_{;e}\, R_{bcf}{}^h{}_{;hgd}
+ R_a{}^e{}_e{}^{f;g}\, R_{bcf}{}^h{}_{;hgd}
- R_a{}^e{}_e{}^{f;g}\, R_{bcg}{}^h{}_{;hfd} ) = 0 .
\end{displaymath}
In terms of indexed invariants and using step 2 this is
\begin{displaymath}
I_{\{0,1,3\},1445} - I_{\{0,1,3\},1451} + I_{\{0,1,3\},2595} = 0 ,
\end{displaymath}
so that $I_{\{0,1,3\},2595}$ is no longer an independent invariant.

\item Commutation of derivatives. Given any tensor
$T^{a_1\ldots a_n}{}_{b_1\ldots b_m}$ we have
\begin{eqnarray*}
\nabla_d \nabla_c T^{a_1\ldots a_n}{}_{b_1\ldots b_m} -
\nabla_c \nabla_d T^{a_1\ldots a_n}{}_{b_1\ldots b_m} =
\qquad\qquad\qquad\qquad \\ =
 \sum_{k=1}^n R_{cde}{}^{a_k}\,T^{a_1\ldots e\ldots a_n}{}_{b_1\ldots b_m}
-\sum_{k=1}^m R_{cdb_k}{}^e\,T^{a_1\ldots a_n}{}_{b_1\ldots e\ldots b_m}.
\end{eqnarray*}
We generate new equations by exchanging the order of all consecutive
covariant derivatives in the canonical invariants after step 1.
As we have already mentioned, in this step different cases are mixed
because second covariant derivatives are converted into Riemann
tensors, and so we find equations involving cases with the same order
$\Lambda$ but larger degree, that is ${\{0,1,3\}}$ and
${\{0,0,1,1\}}$ in our example (\ref{canonical}): under commutation
of $\nabla_g$ and $\nabla_d$ and after the use of steps 1, 2 and 3 we
get
\begin{eqnarray*}
- I_{\{0,1,3\},535} + I_{\{0,1,3\},536} + I_{\{0,1,3\},538}
- I_{\{0,1,3\},539} && \\
+ I_{\{0,1,3\},541} - I_{\{0,1,3\},542} - I_{\{0,1,3\},547}
+ I_{\{0,1,3\},548}  && \\
+ I_{\{0,0,1,1\},385} - I_{\{0,0,1,1\},386} - I_{\{0,0,1,1\},394}
+ I_{\{0,0,1,1\},397} &=& 0 .
\end{eqnarray*}
The equations at this step, once expanded in terms of independent
invariants, become very long, many of them having several thousand terms
for $\Lambda=12$. This is because in step 4 we still have
6368 independent invariants, without even counting the 1639 objects
from the algebraic degree-7 case. The hardest case
to build is \{10\}, taking nearly 300 hours of CPU time and containing
equations with up to 5617 terms (independent invariants).

Our results in columns `Commute' of tables \ref{table12345} and
\ref{table67} perfectly coincide with those of column `Total' of
appendix A of Fulling {\em et al.} \cite{Fulling} after dealing
with the fact that they count product invariants. Note that they do
not study dual invariants and so everything in table \ref{dualtable}
is new.

\item Dimensionally dependent identities (also called Lovelock type
identities): antisymmetrization in $d+1$ indices in dimension $d$
gives zero \cite{Edgar}. We generate a large number of those
equations for dimension 4 by antisymmetrization of random groups of
5 indices in each independent invariant after step 1.

Again, our results in column `4D' of tables \ref{table12345} and
\ref{table67} agree with those of Fulling {\em et al.} \cite{Fulling}.
They give no results for dual invariants.

\item Duals. A product of two $\epsilon$ tensors can be given as a linear
combination of products of $\delta$'s and hence a product of two dual
invariants can be expressed as a linear combination of nondual
invariants. We have combined the dual invariants in pairs in all possible
ways, obtaining 51 relations which allow decomposing 51 independent
invariants after step 5 into products of dual invariants.
\end{enumerate}

We have chosen this order of use of symmetries because only steps 1--5
are signature independent, only steps 1--4 are dimension independent,
steps 1--3 do not mix cases, steps 1--2 do not involve derivatives, and
finally step 1 employs only permutation symmetries, which must be used
first. Note the importance of not decomposing Riemann in its Weyl and
Ricci parts, which would make the process dimension-dependent right
from the outset.

\def\arraystretch{1}
\begin{table}
\begin{center}
\begin{tabular}{|l|ccccccc|}
\hline
Case         & Canon&Invars&Cyclic&Bianchi&Commute& 4D &Duals\\ \hline
\{0\}        & 1    & 1    & 1    & 1     & 1     & 1  & 1   \\ \hline
\{0,0\}      & 4    & 3    & 2    & 2     & 2     & 2  & 2   \\
\{2\}        & 2    & 2    & 2    & 1     & 1     & 1  & 1   \\ \hline
\{0,0,0\}    & 13   & 9    & 5    & 5     & 5     & 3  & 3   \\
\{0,2\}      & 14   & 12   & 9    & 5     & 3     & 3  & 3   \\
\{1,1\}      & 12   & 12   & 9    & 4     & 4     & 4  & 4   \\
\{4\}        & 12   & 12   & 11   & 6     & 1     & 1  & 1   \\ \hline
\{0,0,0,0\}  & 57   & 38   & 15   & 15    & 15    & 4  & 3   \\
\{0,0,2\}    & 119  & 99   & 48   & 27    & 15    & 10 & 10  \\
\{0,1,1\}    & 137  & 125  & 63   & 23    & 23    & 17 & 17  \\
\{0,4\}      & 138  & 126  & 84   & 47    & 3     & 3  & 3   \\
\{1,3\}      & 138  & 138  & 95   & 32    & 5     & 5  & 5   \\
\{2,2\}      & 89   & 86   & 59   & 23    & 7     & 7  & 7   \\
\{6\}        & 105  & 105  & 90   & 50    & 1     & 1  & 1   \\ \hline
\{0,0,0,0,0\}& 288  & 204  & 54   & 54    & 54    & 5  & 3   \\
\{0,0,0,2\}  & 1193 & 1020 & 313  & 175   & 79    & 26 & 25  \\
\{0,0,1,1\}  & 1922 & 1749 & 564  & 194   & 194   & 76 & 74  \\
\{0,0,4\}    & 1647 & 1473 & 648  & 361   & 17    & 12 & 12  \\
\{0,1,3\}    & 3237 & 3099 & 1387 & 442   & 53    & 42 & 42  \\
\{0,2,2\}    & 1735 & 1622 & 727  & 244   & 46    & 34 & 34  \\
\{1,1,2\}    & 1641 & 1617 & 741  & 143   & 67    & 52 & 52  \\
\{0,6\}      & 1770 & 1665 & 1025 & 570   & 3     & 3  & 3   \\
\{1,5\}      & 1770 & 1770 & 1115 & 362   & 5     & 5  & 5   \\
\{2,4\}      & 1770 & 1746 & 1093 & 356   & 9     & 9  & 9   \\
\{3,3\}      & 962  & 962  & 612  & 211   & 9     & 9  & 9   \\
\{8\}        & 1155 & 1155 & 945  & 525   & 1     & 1  & 1   \\ \hline
\end{tabular}
\caption{\label{table12345}%
Number of independent non-dual
invariants after the different steps of simplification: 0) `Canon':
canonical invariants including products of lower degree; 1)
`Invars': canonical invariants without products; 2) `Cyclic':
invariants after imposing the Cyclic symmetry; 3) `Bianchi':
invariants after imposing the Bianchi identity; 4) `Commute':
invariants after commuting covariant derivatives; 5) `4D':
invariants after imposing all possible dimensionally dependent
identities for dimension 4; and 6) `Duals': independent invariants
after decomposition in products of duals. }

\end{center}
\end{table}

\begin{table}
\begin{center}
\begin{tabular}{|l|ccccccc|}
\hline
Case           & Canon &Invars &Cyclic&Bianchi&Commute& 4D &Duals\\ \hline
\{0,0,0,0,0,0\}& 2070  & 1613  & 270  & 270   & 270   & 8  & 4   \\
\{0,0,0,0,2\}  & 14\,408 & 12\,722 & 2495 & 1371  & 549   & 66 & 58  \\
\{0,0,0,1,1\}  & 29\,427 & 27\,022 & 5439 & 1725  & 1725  & 245& 228 \\
\{0,0,0,4\}    & 21\,750 & 19\,617 & 5622 & 3094  & 99    & 37 & 36  \\
\{0,0,1,3\}    & 64\,635 & 60\,984 & 17\,662& 5440  & 577   & 242& 240 \\
\{0,0,2,2\}    & 33\,252 & 30\,974 & 9030 & 2861  & 445   & 169& 165 \\
\{0,1,1,2\}    & 64\,500 & 62\,465 & 18\,272& 3226  & 1235  & 505& 503 \\
\{1,1,1,1\}    & 5684  & 5606  & 1733 & 210   & 210   & 86 & 83  \\
\{0,0,6\}      & 27\,675 & 25\,590 & 10\,600& 5840  & 17    & 12 & 12  \\
\{0,1,5\}      & 54\,930 & 53\,160 & 22\,330& 6938  & 55    & 44 & 44  \\
\{0,2,4\}      & 54\,930 & 52\,764 & 22\,063& 6861  & 93    & 72 & 72  \\
\{1,1,4\}      & 27\,540 & 27\,396 & 11\,695& 2121  & 83    & 66 & 66  \\
\{0,3,3\}      & 27\,986 & 27\,024 & 11\,402& 3597  & 68    & 53 & 53  \\
\{1,2,3\}      & 54\,930 & 54\,654 & 23\,255& 4204  & 212   & 171& 171 \\
\{2,2,2\}      & 9280  & 9104  & 3879 & 715   & 66    & 49 & 49  \\
\{0,8\}        & 26\,670 & 25\,515 & 14\,910& 8225  & 3     & 3  & 3   \\
\{1,7\}        & 26\,670 & 26\,670 & 15\,855& 5000  & 5     & 5  & 5   \\
\{2,6\}        & 26\,670 & 26\,460 & 15\,675& 4950  & 9     & 9  & 9   \\
\{3,5\}        & 26\,670 & 26\,670 & 15\,855& 5000  & 11    & 11 & 11  \\
\{4,4\}        & 13\,685 & 13\,607 & 8111 & 2615  & 12    & 12 & 12  \\
\{10\}         & 15\,120 & 15\,120 & 11\,970& 6615  & 1     & 1  & 1   \\ \hline
\{0,0,0,0,0,0,0\}&19\,610& 16\,532 & 1639 & 1639  & 1639  & 7  & 3   \\ \hline
\end{tabular}
\caption{\label{table67}%
Number of independent invariants for the cases with order $\Lambda=12$,
plus the algebraic case of degree 7. See caption of table
\ref{table12345} for the meaning of the column headers.
}
\end{center}
\end{table}

\begin{table}
\begin{center}
\begin{tabular}{|l|cccccc|}
\hline
Dual case       & Canon &Invars&Cyclic&Bianchi&Commute & 4D \\ \hline
\{0\}$^*$       & 1     & 1    & 0    & 0     & 0      & 0  \\ \hline
\{0,0\}$^*$     & 5     & 4    & 1    & 1     & 1      & 1  \\
\{2\}$^*$       & 3     & 3    & 0    & 0     & 0      & 0  \\ \hline
\{0,0,0\}$^*$   & 35    & 27   & 6    & 6     & 6      & 2  \\
\{0,2\}$^*$     & 63    & 58   & 13   & 5     & 1      & 1  \\
\{1,1\}$^*$     & 36    & 36   & 9    & 2     & 2      & 2  \\
\{4\}$^*$       & 32    & 32   & 11   & 4     & 0      & 0  \\ \hline
\{0,0,0,0\}$^*$ & 288   & 232  & 40   & 40    & 40     & 1  \\
\{0,0,2\}$^*$   & 1059  & 967  & 212  & 98    & 29     & 6  \\
\{0,1,1\}$^*$   & 1095  & 1047 & 236  & 54    & 54     & 13 \\
\{0,4\}$^*$     & 920   & 876  & 285  & 128   & 1      & 1  \\
\{1,3\}$^*$     & 920   & 920  & 296  & 60    & 2      & 2  \\
\{2,2\}$^*$     & 484   & 478  & 163  & 37    & 3      & 3  \\
\{6\}$^*$       & 435   & 435  & 220  & 95    & 0      & 0  \\ \hline
\{0,0,0,0,0\}$^*$& 3031 & 2582 & 330  & 330   & 330    & 2  \\ \hline
\end{tabular}
\caption{\label{dualtable}%
Number of independent dual invariants after the different steps of
simplification. See the caption of table \ref{table12345} for the meaning
of the column headers. Note that there is no final column `Duals' in this
table.
}
\end{center}
\end{table}

\section{Implementation}
The implementation of the new database of relations among the
differential invariants closely follows that of our previous version
of {\em Invar} \cite{Invar}. In particular the commands given in
appendices C and D of \cite{Invar} are still valid, with minimal
changes, the most important of them being the fact that now the
different cases are identified by the case list of orders
\texttt{\{0,1,3\}} in the {\em Mathematica} version and \texttt{[0,1,3]} in
the {\em Maple} version rather than an integer degree. Therefore an
invariant
is now denoted as \texttt{RInv[\{0,1,3\}, 2595]} or
\texttt{DualRInv[\{1,3\}, 920]} in {\em Mathematica} and
\texttt{RInv[[0,1,3], 2595]} or \texttt{DualRInv[[1,3], 920]} in
{\em Maple}. The old notation \texttt{RInv[3, 9]} for an algebraic
invariant is automatically translated to its new notation
\texttt{RInv[\{0,0,0\}, 9]} in {\em Mathematica} and \texttt{RInv[[0,0,0],
9]} in {\em Maple}.

The system now handles derivatives of tensorial expressions.
The covariant derivative $\nabla_a$ is treated as an operator and denoted
\texttt{CD[-a]} in our tensor computer algebra systems. Hence, we can
represent the invariant
\begin{displaymath}
I_{\{1,1\},2} \equiv R^{ab}{}_{ab}{}^{;c}\; R_c{}^d{}_d{}^e{}_{;e}
= - R^{ab}{}_{;a} R_{;b}
\end{displaymath}
as \texttt{CD[c][R[a,b,-a,-b]]*CD[-e][R[-c,d,-c,e]]} in {\em Mathematica},
and as \texttt{CD[c](R[a,b,-a,-b])*CD[-e](R[-c,d,-c,e])} in {\em Maple}.

Covariant derivatives in {\em xTensor} and {\em Canon} are linear operators
and automatically obey the standard rules \cite{Wald}, in particular the
Leibnitz rule for products.
The canonicalization algorithms consistently manipulate derivatives and
only one comment is required, concerning how the different Riemann tensors
are sorted in a given invariant: we follow the convention of having tensors
with more covariant derivatives on the right to help the canonicalizer,
which starts placing indices on the left. That is why we chose
the convention $\lambda_i\le \lambda_{i+1}$ for the cases
$\{\lambda_1,\ldots,\lambda_n\}$, opposite to the choice by
Fulling {\em et al.} \cite{Fulling}.

As a simple example of use of the main command \texttt{RiemannSimplify}
we can check the interesting relation
\begin{displaymath}
R^{abcd;e}{}_a \, R_{be}{}^{fg;hi} \, R_{cfgi;dh} =
\frac{1}{8} \; R^{abcd;e}{}_e\, R_{ab}{}^{fg;hi}\, R_{cdfg;ih} ,
\end{displaymath}
in which the rearrangement of the lower indices produces a factor $1/8$.
This is rule number 2868 in the file of Bianchi relations among the
invariants of case \{2,2,2\}. We can check it in {\em Mathematica} using
the sentence
\begin{verbatim}
  expr = CD[-a]@CD[e]@R[a,b,c,d] * CD[i]@CD[h]@R[-b,-e,f,g]
             * CD[-h]@CD[-d]@R[-c,-f,-g,-i]
       - 1/8 * CD[-e]@CD[e]@R[a,b,c,d] * CD[i]@CD[h]@R[-a,-b,f,g]
             * CD[-h]@CD[-i]@R[-c,-d,-f,-g];
  RiemannSimplify[ expr ]
\end{verbatim}
\[ 0 \]\vspace*{-0.7cm}
where \texttt{f@x} is a {\em Mathematica} shorthand for \texttt{f[x]}.
The corresponding {\em Maple} expression can be obtained replacing each
\texttt{f@x} by \texttt{f(x)}. The final call would be
\begin{verbatim}
  RiemannSimplify( expr )
\end{verbatim}
\[ 0 \]

The database of relations is arranged as follows: there is a file for
each step and case (so $6\times 48 + 5\times 15 = 363 $ files in total)
containing the relations at that step among the invariants of that case
and all previous cases. The full database takes more than 1.5 Gbytes of
memory and even more once it is read by the computer algebra systems.
Most of it comes from the 13 last cases of order 12 at the commutation
step (step 4), and so we provide alternative smaller files in which the
rules have not been fully expanded in terms of independent invariants
of previous cases. The user can configure the system to read either
the fully expanded files or the non-expanded files in those cases. If
the latter are chosen then repeated (up to four times) use of the
simplification functions might be needed to fully expand an invariant,
resulting in a slower process. With non-expanded files we have 365Mbytes
of data.

\section{Conclusions}
\vspace{-3mm}
We have successfully extended our tensor computer algebra package
{\em Invar} to handle differential invariants of the Riemann tensor up to
12 derivatives of the metric. Unlike the problem of algebraic
invariants, to which a great deal of effort has been dedicated in
the last decades, the problem of differential invariants remained
almost unexplored due to its much more difficult character: two new
sources of equations come into play, namely the Bianchi identity and
the non-commutation of covariant derivatives. This article has
completely solved the problem for those cases most likely to be
needed in current computations. The problem has been solved in the
most useful way: by providing a database with all equations coming
from multiterm symmetries, such that any invariant can be
immediately looked up without the need of inefficient intermediate
computations.

The new database is much larger than its previous version, now containing
645\,625 relations for 647\,817 canonical invariants (counting both
dual and nondual invariants). This represents an increase by a factor
larger than 30 with respect to the 21\,221 relations for 21\,246
invariants in the old database. But this investigation is not only a
quantitative extension of \cite{Invar}. We have also extended our
algorithms to handle derivatives and their associated symmetries: Bianchi
and commutation. The latter has been especially hard, producing equations
with thousands of terms, such that the database has increased in size
by a factor larger than 250 with respect to the previous version.

{\em Invar} runs on top of the multipurpose tensor computer algebra systems
{\em xTensor} \cite{xTensor} for {\em Mathematica} and
{\em Canon} \cite{Canon} for {\em Maple}.

A final step is required to construct a computer algebra system for
generic treatment of the Riemann tensor: manipulation of expression with
free indices. We are currently analyzing this extension.

\section*{Acknowledgements}
JMM thanks Brian Edgar for helpful discussions and the University of
Link\"oping for hospitality.
JMM was supported by the Spanish MEC under the research project
FIS2005-05736-C03-02. DY acknowledges the grant ``Beca de
introducci\'on a la investigaci\'on'' from CSIC. RP acknowledges
grant no. 2898-07-1 from CAPES. Part of the
computations were performed at the {\em Centro de SuperComputaci\'on
de Galicia} (CESGA).

\appendix

\end{document}